\documentclass{aip-cp}

\usepackage[numbers]{natbib}
\usepackage{rotating}
\usepackage{graphicx}

\begin{document}

% Title portion
\title{Golden Probe of the Top Yukawa}

\author[aff1]{Yi Chen}
\author[aff2]{Daniel Stolarski\corref{cor1}}
\author[aff3]{Roberto Vega-Morales}
%\eaddress[url]{http://www.aip.org}
%\author[aff2,aff3]{Author's Name}
%\eaddress{anotherauthor@thisaddress.yyy}

\affil[aff1]{Lauritsen Laboratory for High Energy Physics, California Institute of Technology, Pasadena, CA, USA}
\affil[aff2]{Theory Division, Physics Department, CERN, CH-1211 Geneva 23, Switzerland}
\affil[aff3]{CAFPE and Departamento de F'sica Te—rica y del Cosmos, Universidad de Granada, E-18071 Granada, Spain}
\corresp[cor1]{Corresponding author: daniel.stolarski@cern.ch}

\begin{flushright}  
{\normalsize{  
CAFPE 188/15, CERN-PH-TH-2015-301, UG-FT 318/15
  }   
  }
\end{flushright}
\vspace*{-1.5cm}  
{\let\newpage\relax\maketitle}
%\maketitle

\begin{abstract}
We describe how the Higgs decay to four leptons can be used to probe the nature and $CP$ structure of the top Yukawa coupling. 
\end{abstract}

\section{HIGGS DECAY TO FOUR LEPTONS}

The discovery of the Higgs is the beginning of a long program to study its properties. Current state of the art characterizations of the boson usually involve looking at partial rates: comparing the frequency of different production and decay rates to the Standard Model (SM) prediction. In the Higgs decay to $4\ell$ ($4e,\,4\mu,\,2e2\mu)$, there is significantly more information than just how often this decay happens. Assuming that the Higgs is a scalar, there are five kinematic variables that describe each event. These can be parameterized, for example, using angles between a lepton and $Z$ momentum, and the invariant mass of lepton pairs. 

Shortly after the discovery, it was shown~\cite{Stolarski:2012ps} that these kinematic variables encode information about the tensor structure of the Higgs coupling to gauge bosons. We compared three different possibilities:
\begin{equation}
h \,Z^\mu Z_\mu \;\;\; {\rm or} \;\;\; h\, Z^{\mu\nu} Z_{\mu\nu} \;\;\; {\rm or} \;\;\; h\, Z^{\mu\nu} F_{\mu\nu}
\end{equation}
where $h$ is the putative Higgs, $Z_{\mu\nu}=\partial_\mu Z_\nu-\partial_\nu Z_\mu$, and $F_{\mu\nu}$ is the equivalent field strength for the photon. We showed that with $\mathcal{O}(50)$ events at the LHC, these different possibilities can be distinguished.

\section{PROBING LOOP PROCCESSES}

\begin{figure}[h]
  \centerline{\includegraphics[width=420pt]{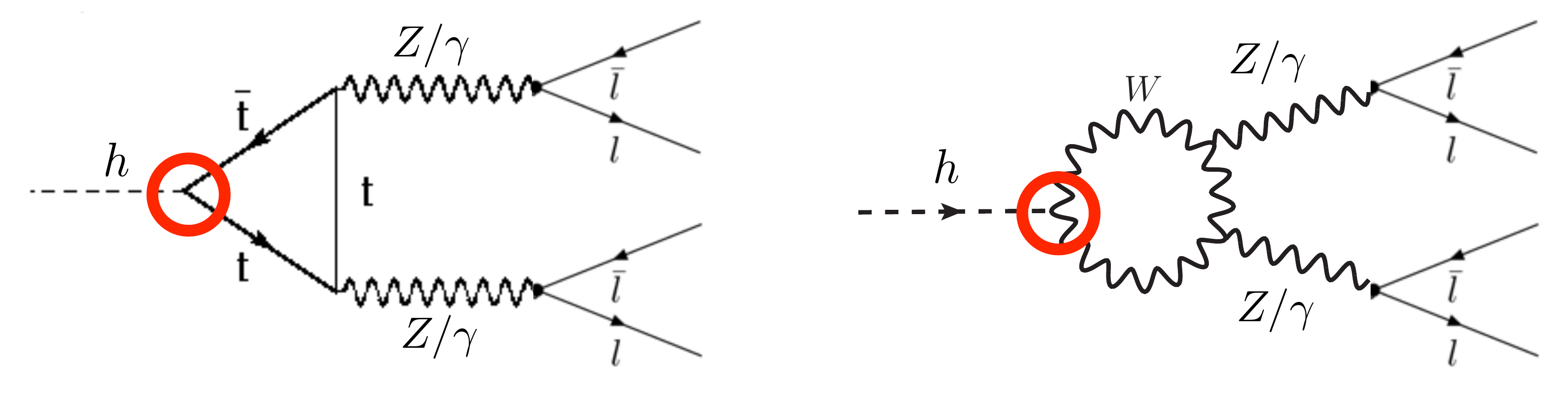}}
  \caption{Feynman diagrams for the leading NLO corrections to $h\rightarrow4\ell$. The coupling of the Higgs is circled to show which Higgs coupling can be probed, namely those to the top and $W$. }
  \label{fig:Feyn}
\end{figure}

Having seen that kinematic distributions in $h\rightarrow4\ell$ can be useful, we now turn our attention to measuring loop processes in this channel, namely the next-to-leading order (NLO) corrections to the tree level effect induced by the $h \,Z^\mu Z_\mu$ operator. The largest of these are shown in Figure~\ref{fig:Feyn}, showing that this channel can be sensitive to the couplings of the Higgs to the top and $W$. Here we examine the sensitivity of this channel to the top Yukawa coupling, so we keep all other couplings (Higgs to $W$ and $Z$ and gauge boson couplings to fermions) at their SM values.

% Figure

We parameterize the top Yukawa coupling as
\begin{equation}
h \, \bar{t}\, \left(y+i\,\tilde{y}\,\gamma^5 \right) t\;.
\end{equation}
In the SM, $y\approx 1$ and $\tilde{y}\approx 0$. The pseudo-scalar operator is $P$ and $CP$ odd, so if both $y$ and $\tilde{y}$ are non-zero, then $CP$ is violated in the top Yukawa coupling. This effect is tiny in the SM, so a detection of this would be a clear sign of new physics. 

\subsection{Other Probes of $CP$ Violation}

If there is $CP$ violation in the top Yukawa coupling, this contributes to the electric dipole moments (EDM) of the electron, neutron, and Hg atom~\cite{Brod:2013cka} at two loops. These bounds, particularly the electron EDM, constrain $\tilde{y}$ to be less than $\mathcal{O}(1\%)$, with future experiments expected to reduce the bound to one in ten thousand. 

The computations of EDM bounds assume SM couplings for other fields. In particular, the Yukawa coupling of the first generation fermions to the Higgs play a key role. As we have no direct experimental evidence that these couplings are SM-like, one could also consider the scenario where those couplings are zero, and in that case $\mathcal{O}(1)$ values of $\tilde{y}$ are allowed. In this case, the neutron EDM is still somewhat sensitive due to the Weinberg operator, and future experiments are expected to bound $\tilde{y}$ at the per mille level, but they could also see a discovery. In that case, direct measurements of the top Yukawa coupling will be critical to characterizing the nature of $CP$ violation responsible.

\subsection{Experimental Sensitivity}

\begin{figure}[h]
  \centerline{\includegraphics[width=190pt]{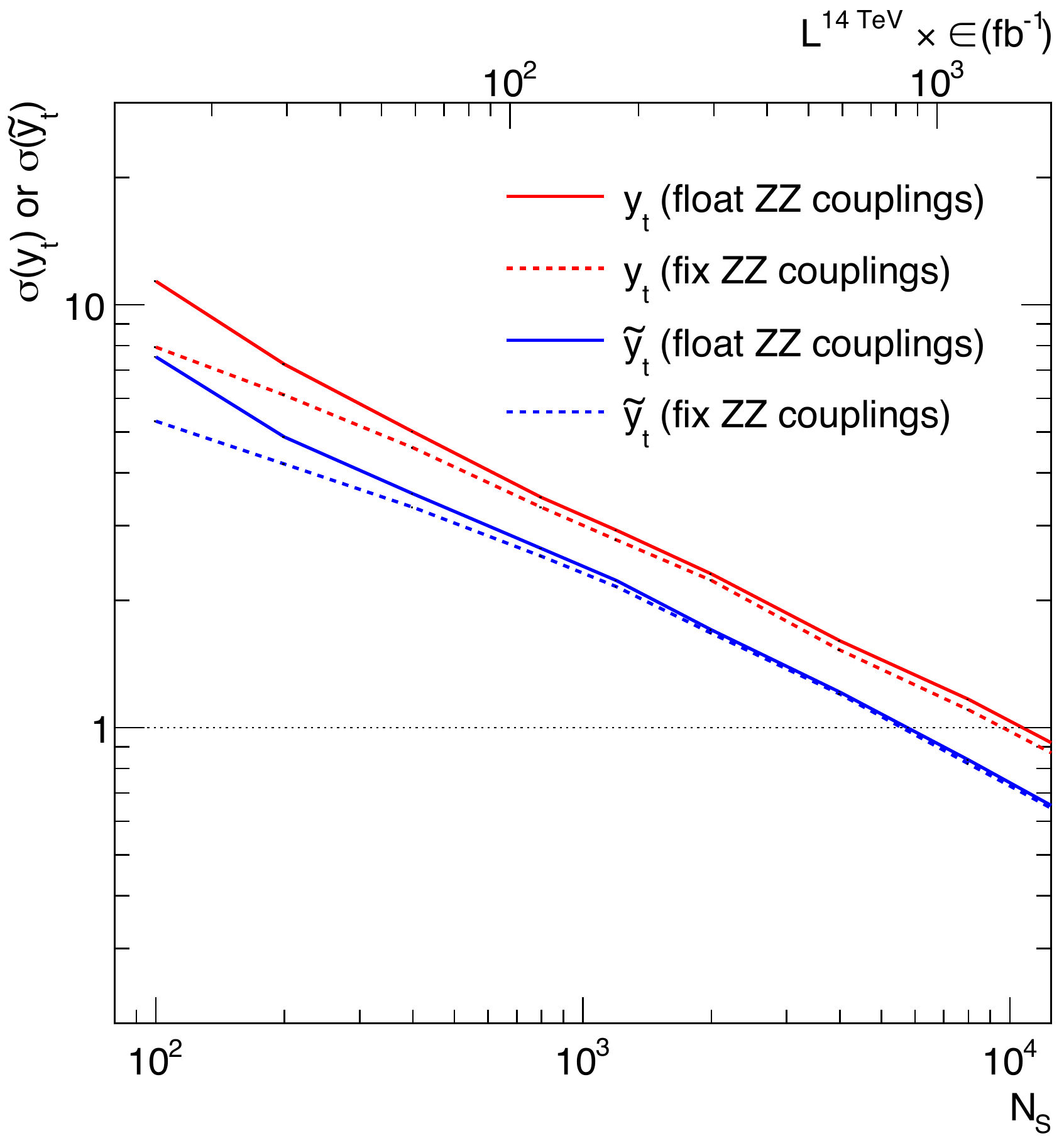}}
  \caption{Sensitivity of the measurements described here in to $y$ and $\tilde{y}$ as a function of the number of events. For details see text and~\cite{Chen:2015rha}. }
  \label{fig:nev}
\end{figure}

In~\cite{Chen:2015rha} we analyzed the experimental sensitivity of $h\rightarrow4\ell$ to $y$ and $\tilde{y}$. The sensitivity depends on the number of such decays which in turn depends on the integrated luminosity. To get an $\mathcal{O}(1)$ precision on $y$ and $\tilde{y}$, one needs about ten thousand events, which corresponds to a few thousand fb$^{-1}$ of luminosity and depends on the experimental efficiencies. The scaling of the sensitivity is shown in Figure~\ref{fig:nev}. From there, we see that the sensitivity to $\tilde{y}$ is better than to $y$, and this is because there is no SM contribution of the $W$ to compete with in the $P$ odd channel.

% Figure

The sensitivity depends on the experimental cuts used to collect $h\rightarrow4\ell$. Current cuts are designed for discovery and to maximize the ratio of signal to background. As an example, they require one of the lepton pairs to have invariant mass bigger than 40 GeV at CMS. Since we already have the discovery, current measurements would be more sensitive to NLO effects if they aimed to maximize signal efficiency~\cite{Chen:2015iha}, even if that means having more background events in the sample. In particular, by loosening the invariant mass cut down to 4 GeV, the experiments will be much more sensitive to effects with photon intermediate states. This will increase the amount of background, but~\cite{Chen:2015iha} showed that the sensitivity to loop effects is much improved even with background taken into account. 

We can plot the the $1-\sigma$ contours in the $y-\tilde{y}$ plane for different numbers of events, and this is done Figures~\ref{fig:800}, \ref{fig:8000}, and~{\ref{fig:20000}. All our simulations use a crude modelling of detector effects including energy smearing described in~\cite{Chen:2015iha}. We show three different contours for different assumptions about how events are collected.  The outermost is using current CMS cuts, the middle is using the realistic ``Relaxed-$\Upsilon$'' cuts described in~\cite{Chen:2015iha}, and the innermost is with zero background, the theoretically best possible result. We see that definite improvements can be made relative to current cuts, but it is possible that a still better method exists. 

% Figure
\begin{figure}[h]
  \centerline{\includegraphics[width=190pt]{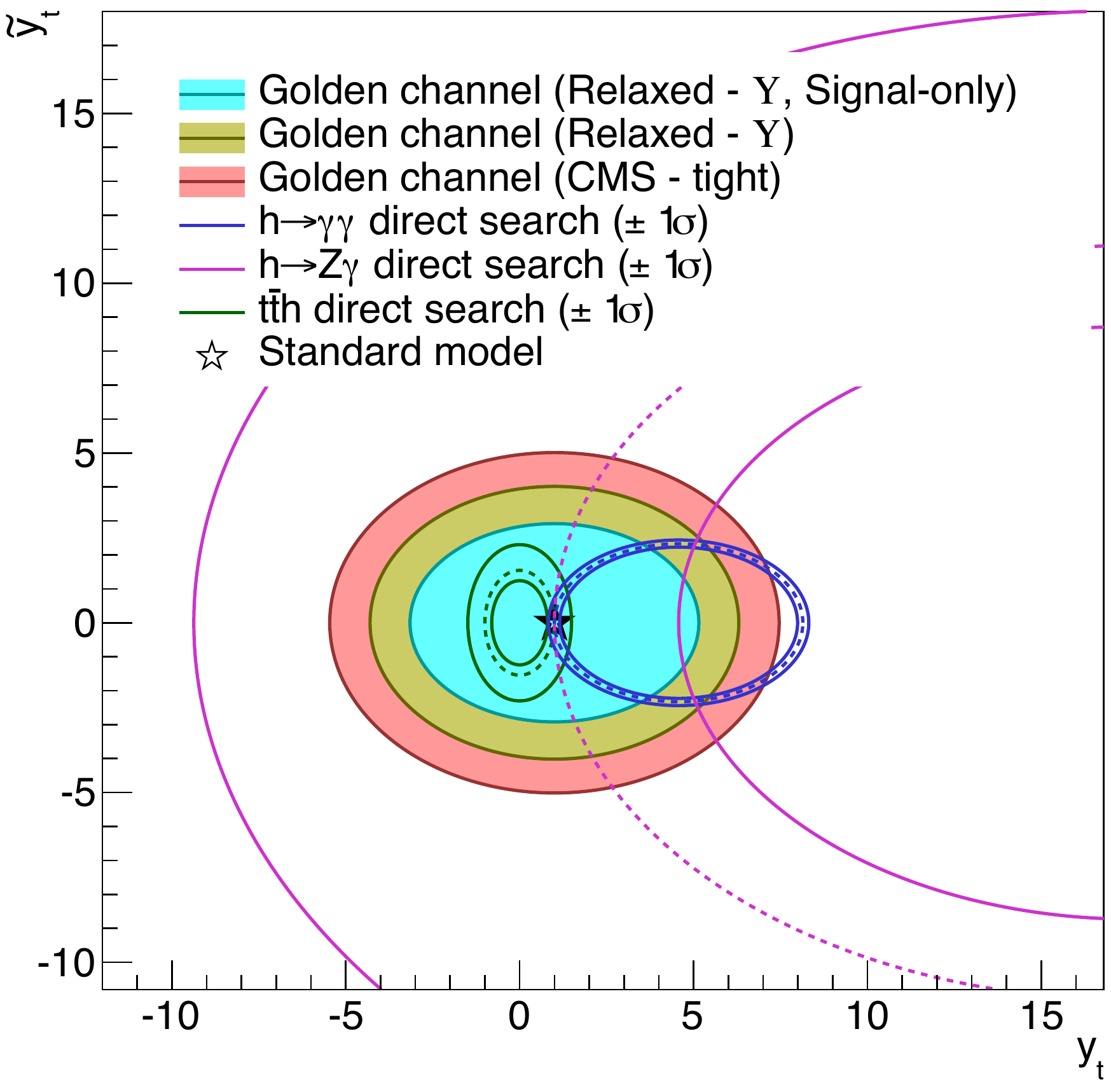}}
  \caption{Sensitivity of the measurements described here in the $y-\tilde{y}$ plane with 800 events corresponding to approximately 300 fb$^{-1}$. For details see text and~\cite{Chen:2015rha}. }
  \label{fig:800}
\end{figure}

% Figure
\begin{figure}[h]
  \centerline{\includegraphics[width=190pt]{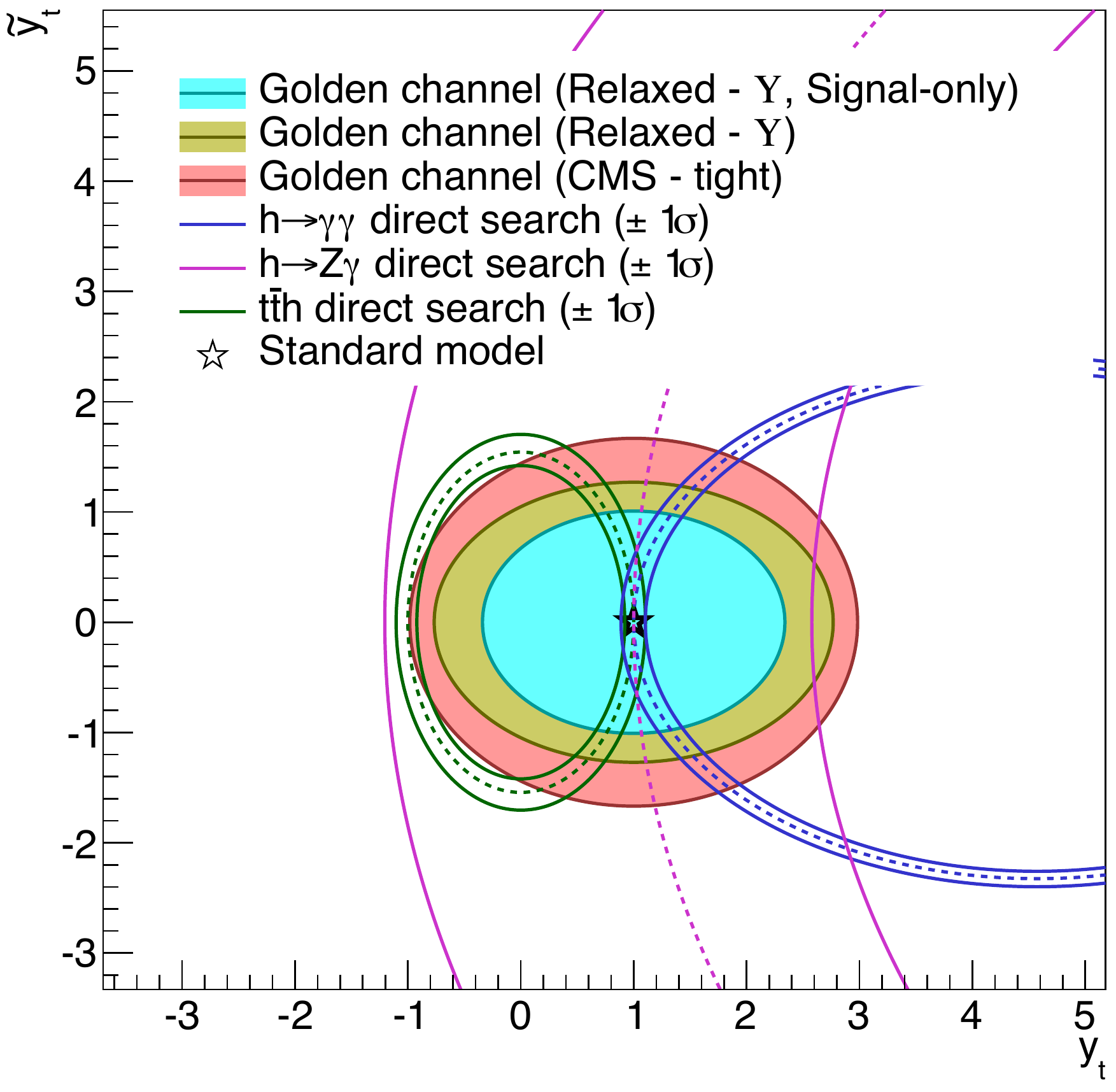}}
  \caption{Same as Figure~\ref{fig:800} but for 8,000 events corresponding to approximately 3,000 fb$^{-1}$.}
  \label{fig:8000}
\end{figure}

% Figure
\begin{figure}[h]
  \centerline{\includegraphics[width=190pt]{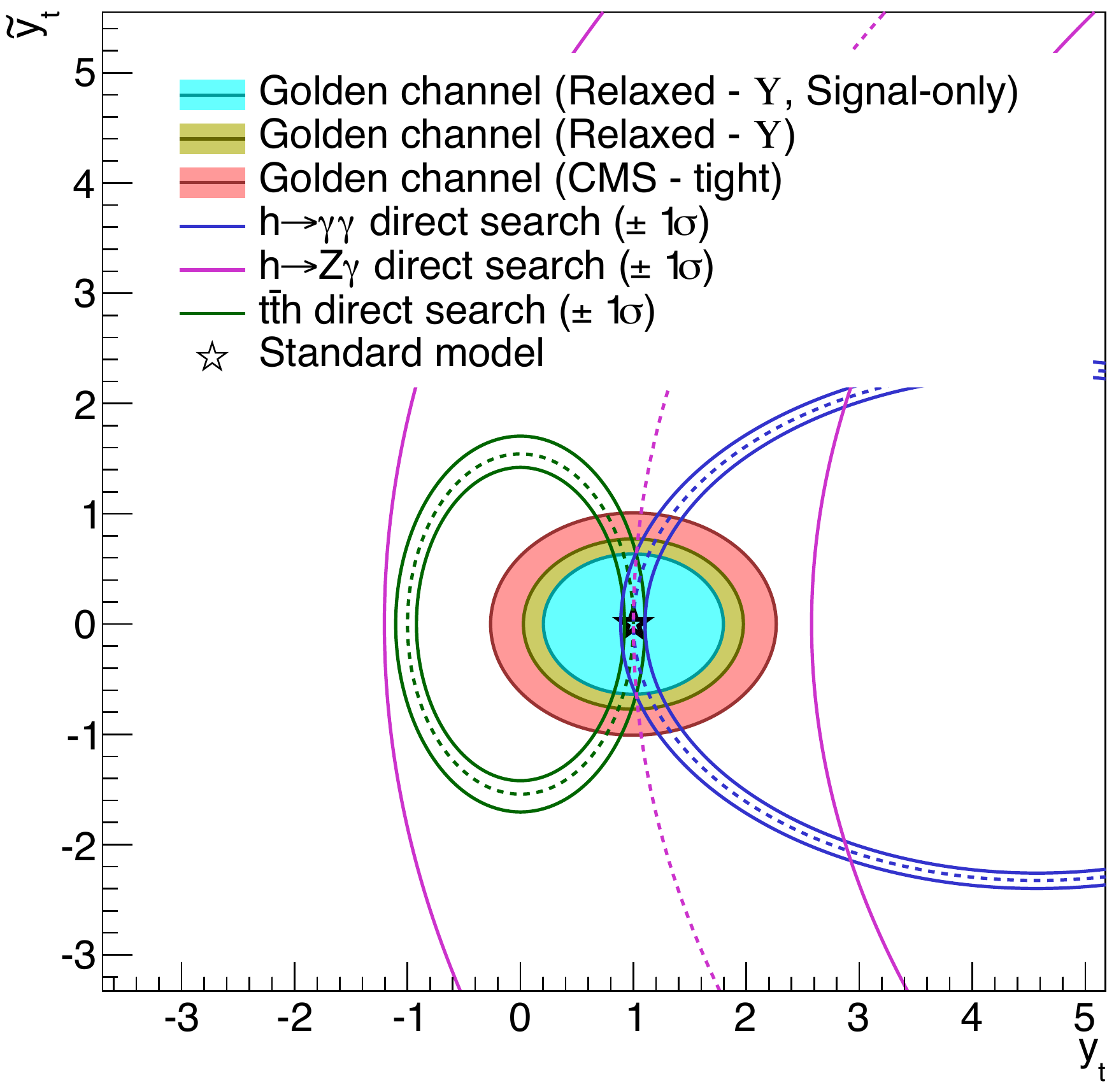}}
  \caption{Same as Figure~\ref{fig:800} but for 20,000 events corresponding to a putative 100 TeV collider with 3,000 fb$^{-1}$.}
  \label{fig:20000}
\end{figure}

There are other measurements that are sensitive to the top Yukawa coupling, and we have put some of them in the figures for comparison. The green oval is the bound from $t\bar{t}h$ production which is quadratically sensitive to $y$ and $\tilde{y}$. The thickness of the oval represents putative uncertainty of that measurement, but even with an infinitely precise measurement, $t\bar{t}h$ alone will never be able to determine at what point in the oval the true theory lies. The blue band is $h\rightarrow \gamma\gamma$, which is again an oval, but displaced from the origin because of the contribution from the $W$ loop to that process. The pink curve is $h\rightarrow Z\gamma$ which is much less sensitive.

In Figure~\ref{fig:800}, we show the projected sensitivity with 800 events which corresponds to roughly 300 fb$^{-1}$, where we see the $h\rightarrow 4\ell$ measurement is only barely competitive with the others. On the other hand, with higher luminosity the other measurements do not gain much sensitivity, while $h\rightarrow 4\ell$ will always be statistics limited. Therefore we see in Figure~\ref{fig:8000}, which corresponds to roughly 3,000 fb$^{-1}$, the high luminosity run of the LHC, that this measurement gets substantially better, allowing for strong constraints. Finally, we see that with 20,000 events, the measurement is even further improved, and this quantity of events could be achieved with a 100 TeV collider recording 3 ab$^{-1}$.

If $t\bar{t}h$ and $h\rightarrow \gamma\gamma$ are both measured to be SM-like in the asymptotic future of the LHC, then they can mutually break each others degeneracy in $y-\tilde{y}$ plane. On the other hand, there are assumptions about no other new physics that have to go into this measurement, particularly in the loop induced $h\rightarrow \gamma\gamma$ decay. Furthermore, if there is a deviation from the SM prediction in one or both of these measurements, then the $h\rightarrow 4\ell$ analysis described here will be crucial in characterizing the deviation.

\section{CONCLUSIONS}

The four lepton decay of the Higgs is an excellent channel to make detailed measurements. The four body final state gives rise to a rich structure of kinematic variables that can be exploited to measure the properties of the Higgs. This analysis is complementary to rate measurements and can give information not available in other ways. In particular, this channel is sensitive to NLO effects that interfere with the tree level contribution, giving access to the Higgs' coupling to the top quark and $W$ boson. Furthermore, because the NLO effects interfere with the leading order, one can measure signs and phases of the Yukawa coupling. Therefore, this measurement can be used to place model independent bounds, or possibly even discover new physics.

% Sections that will go in second font

% Acknowledgement
\section{ACKNOWLEDGMENTS}
We would like to thank the organizers of LHCP and the top sessions for the invitation to give this talk. RVM is supported by MINECO, under grant number FPA2013-47836-C3-2-P. 

\pagebreak

% References

%\nocite{*}
%\bibliographystyle{aipnum-cp}%
%\bibliographystyle{apsrev}
\bibliographystyle{JHEP}
\bibliography{h4l}%

\end{document}